# Improving Freeway Merging Efficiency via Flow-Level Coordination of Connected and Autonomous Vehicles

Jie Zhu, Ivana Tasic, and Xiaobo Qu, *Member, IEEE*

**Abstract—Freeway on-ramps are typical bottlenecks in the freeway network due to the frequent disturbances caused by their associated merging, weaving, and lane-changing behaviors. With real-time communication and precise motion control, Connected and Autonomous Vehicles (CAVs) provide an opportunity to substantially enhance the traffic operational performance of on-ramp bottlenecks. In this paper, we propose an upper-level control strategy to coordinate the two traffic streams at on-ramp merging through proactive gap creation and platoon formation. The coordination consists of three components: (1) mainline vehicles proactively decelerate to create large merging gaps; (2) ramp vehicles form platoons before entering the main road; (3) the gaps created on the main road and the platoons formed on the ramp are coordinated with each other in terms of size, speed, and arrival time. The coordination is formulated as a constrained optimization problem, incorporating both macroscopic and microscopic traffic flow models, for flow-level efficiency gains. The model uses traffic state parameters as inputs and determines the optimal coordination plan adaptive to real-time traffic conditions. The benefits of the proposed coordination are demonstrated through an illustrative case study. Results show that the coordination is compatible with real-world implementation and can substantially improve the overall efficiency of on-ramp merging, especially under high traffic volume conditions, where recurrent traffic congestion is prevented, and merging throughput increased.**

*Index Terms*—connected and autonomous vehicles, coordinative merging strategy, freeway on-ramp merging, microscopic simulation, optimization.

## I. INTRODUCTION

### A. Background and Objective

ON-RAMP merging areas are typical bottlenecks in the freeway network, where the lane-changing maneuvers of merging vehicles impose frequent disturbances on the traffic flow and cause reduced operational efficiency and a high risk of traffic breakdown and capacity drop [1]. Improving traffic operation at on-ramps will benefit traffic in the entire freeway network, and thus has been of utmost importance in the continuous research efforts. Prior efforts to facilitate on-ramp merging include ramp metering systems [2-6], variable speed limits [7-10], and the combination of both [11-14]. Though these approaches can improve the ramp merging operation, the improvements are somehow limited because the control only takes place at an aggregated level.

The emerging Connected and Autonomous Vehicles (CAVs) hold the potential to regulate individual vehicles, presenting an opportunity to control traffic at a disaggregated level [15]. The communication capability enables cooperative driving by allowing for detailed information exchange among road users and infrastructures [16]. Such cooperation is further facilitated by the precise and timely control of vehicle dynamics enabled by the autonomous driving systems. Based on these emerging vehicle technologies, many studies are devoted to exploring the cooperation possibilities and potential impacts of CAVs in various traffic bottlenecks [17-20]. Most of these studies expect improved traffic operation with the presence of well-designed control and sufficient market penetration of CAVs.

The potential to promote on-ramp merging via CAVs is also discussed in the literature. A common practice is to regulate the interactive behaviors between a ramp merging vehicle and its direct neighbors for efficiency and/or safety benefits. For example, [21] maps a pair of competing mainline and ramp vehicles to each other's lane as virtual vehicles for collision-free merging. References [22] and [23] jointly design the trajectories of a competing pair of vehicles through optimization for efficiency improvements. Later, [24] further adapts [23] for restrained negative safety impacts on the following traffic. References [25] and [26] also investigate the safety impacts of CAV merging and suggest that CAVs, especially with predictive control strategies, can reduce the frequency and severity of merging conflicts. More recently, [27] describes the merging process as a two-player dynamic game, where each vehicle makes trajectory decisions to maximize its own driving utility, while considering the potential actions of the competing vehicle. [28] investigates cooperation under mixed traffic with CAVs and Human-Driven Vehicles (HDVs), where the authors distinguish different combinations of CAVs and HDVs in a merging triplet (i.e., a merging vehicle and its putative leader and follower) and design for each combination a cooperative strategy that checks the desired distance and

Manuscript is submitted for review in July 2021. This work was funded by the Area of Advance Transport at Chalmers University of Technology. *(Corresponding author: Xiaobo Qu)*

Jie Zhu, Ivana Tasic, and Xiaobo Qu are with the Department of Architecture and Civil Engineering, Chalmers University of Technology, 41296 Gothenburg, Sweden. (email: jie.zhu@chalmers.se; tasic@chalmers.se; drxiaoboqu@gmail.com).

speed at a series of set-points. Although these approaches can facilitate smooth merging at on-ramps, they focus on the interaction between individual vehicles, whereas the superiority in a continuous traffic flow are not always guaranteed.

Another branch of CAV merging strategies extends the scope of control from single pair/triplet of vehicles to multiple vehicles within a communication range. By assuming the presence of an upper-level merging sequence, many studies are devoted to formulating the lower-level trajectories of relevant vehicles under an optimization framework. The established models target at different objectives favoring traffic efficiency [29-34], energy use [35-38], or passenger comfort [35], while being subject to vehicle dynamics and safety requirements. An alternative method is provided in [39], where a central controller chooses from alternative actions (i.e. speed up, slow down, change lanes) of a ramp vehicle and its mainline competitors by comparing the total travel time resulted from each option. Despite the differences in assumptions and methods, all these studies focus on the lower-level operational control of CAVs, whereas the upper-level decisions are either totally ignored or considered in a very simple manner (e.g., assuming a first-in-first-out merging sequence). This leaves a chance to further enhance merging coordination through more efficient upper-level control.

A few recent works shed light on the upper-level decisions. For example, the merging sequence can be locally adjusted through predetermined rules [40], optimization criteria [41], or generic algorithms [42], in order to improve collective benefits in efficiency, energy use, and/or passenger comfort. Alternatively, some other works determine the target gap for each ramp vehicle based on the trajectory costs of leading the vehicle into different gaps [43], [44]. These methods can improve the upper-level efficiency of ramp merging; however, they only consider one ramp vehicle at a time, so the efficiency of other ramp vehicles and the option of group merging are disregarded. In [45], a flow-level merging strategy is proposed, where the mainline traffic is periodically compacted to create large gaps, and the ramp vehicles are released into the gaps through ramp metering signals. However, this strategy stipulates that the release of ramp vehicles fully depends on the mainline conditions, so the efficiency of ramp traffic is not actively considered. Recently, [46] adopt the similar idea of periodic gap creation and combine it with a batch merging strategy to close the extra time gaps induced by the lane-changing maneuvers. The benefits of the proposed system are demonstrated in theory, but no numerical/simulation experiment is carried out. In addition, both [45] and [46] focus more on the definition and validation of the proposed systems, without discussing how to maximize the expected coordination benefits with respect to the real-time traffic conditions.

In summary, the existing CAV ramp merging strategies mainly deals with the lower, local-level control, such as the trajectory decisions of individual vehicles, whereas the upper, flow-level control options are discussed to a very limited extent. In addition, most existing strategies tend to merge the ramp vehicles one-by-one without exploring the benefit to manipulate traffic flow and guide multiple ramp vehicles into a single gap. The very few efforts on flow-level coordination focus on the theoretical validation, rather than the practical implementation under real-time traffic operation. In this study, we seek to partially make up the above research gaps by developing a novel upper-level merging strategy that coordinates the two streams of traffic (instead of individual vehicles) for maximized flow-level gains in ramp merging efficiency and traffic flow stability.

### B. Research Approach and Contributions

In this study, a CAV-enabled ramp merging strategy, called Coordinative Merging Control (CoMC), is proposed to facilitate efficiency and stability at freeway on-ramps. The strategy combines two promising ideas: (1) proactive creation of large gaps on the main road, and (2) platooning of ramp merging vehicles. Gap creation is a primary way to promote on-ramp merging. Common practices include the reservation or slot-based methods [21], [47], trajectory planning methods [23], [29], [30], [35], [36], and proactive gap creation systems [45], [46]. Previous results show that gap creation can eliminate critical merging situations by providing ramp vehicles with readily available gaps. Platoon driving is another approach that is proven to stabilize traffic flow and increase throughput in various bottlenecks [48-50]; however, it is rarely applied to the merging of on-ramp vehicles. In this research, we integrate the ideas of gap creation and platoon merging in a coordination framework, under which the mainline traffic is compacted to create large gaps when triggered by the formation of the ramp merging platoons, and the merging platoons are smoothly directed into the created gaps through the control of speed and arrival time. The coordination is formulated as an optimization problem to determine the optimal control plan adaptive to real-time traffic conditions. The model incorporates microscopic and macroscopic traffic flow models to account for the dynamics of individual vehicles, as well as the flow-level stability and efficiency gains. The benefits of the proposed CoMC strategy are demonstrated through an illustrative case study conducted on a microscopic simulation platform. Results show clear efficiency gains of the CoMC strategy, especially under high traffic volume conditions.

Table I compares the CoMC strategy with the existing CAV ramp merging approaches and highlights its novelty:

- The CoMC strategy controls the two streams of traffic (instead of individual vehicles) for the flow-level efficiency gains. We expect that the consolidation of gap creation and platoon merging can result in enhanced coordination benefits than applying them separately.
- This strategy, for the first time to the authors best knowledge, explicitly considers the platooning of ramp merging vehicles and determines the optimal platoon formation with respect to traffic conditions.
- This strategy incorporates the macroscopic traffic flow models, which allow for an explicit consideration on the transition of the fundamental traffic state and the stability of the traffic flow.

The remaining of this paper is structured as follows. Section II introduces the CoMC strategy and presents its analytical formulation in an optimization framework. Section III provides a solution method of the optimization model. Section IV presents a case study and discusses the efficiency of CoMC under various demand scenarios. The conclusion is drawn in Section V.

## II. COORDINATIVE MERGING CONTROL

### A. Coordinative Merging Control (CoMC) Strategy

The CoMC strategy, combining proactive gap creation and platoon merging, consists of three control components: (1) *mainline control*: mainline vehicles decelerate in advance to create large gaps on the main road; (2) *ramp control*: merging vehicles form platoons on the on-ramp ; (3) *centralized coordination*: the gaps created on the main road and the platoons formed on the ramp are coordinated by a control center in terms of size, speed, and arrival time. As illustrated in Fig. 1, the coordination is carried out in the following steps:

**Step 1**: Upon arrival, the ramp vehicles stop at a pre-specified position on the ramp and register themselves with the control center.

**Step 2**: The control center counts the number of ramp vehicles arriving. When a certain number of ramp vehicles has accumulated, the control center initiates coordinative merging by appointing a mainline vehicle as the facilitating vehicle and sending instructions on where and how much this vehicle should cooperatively decelerate.

**Step 3**: The facilitating vehicle accepts the cooperation request (otherwise the request is passed to the next vehicle) and sends back a confirmation to the control center. Then, it executes the required deceleration and develop a gap from its original leader.

**Step 4**: The control center releases the vehicles waiting on the ramp as a platoon by specifying their moving trajectories.

**Step 5**: The ramp vehicles follow the instructions from the control center when driving towards the merging point.

In order to achieve smooth and efficient merging, the mainline cooperation and the ramp platoon formation should be coordinated in terms of three requirements: (1) the created mainline gap should be large enough for the platoon to merge into (the requirement of size); (2) the platoon should reach the same speed as the mainline facilitating vehicle when arriving at the merging point (the requirement of speed); (3) the gap should be just available at the merging point when the platoon arrives there (the requirement of arrival time).

Fig. 1. Coordinative Merging Control (CoMC) system

Note that, the essence of CoMC is to collect space on the main road by compacting mainline vehicles and to make full use of the collected space by grouping merging vehicles into proper platoons. This can be explained by macroscopic traffic flow theories [51], [52]. Fig. 2 shows a generalized fundamental diagram, where each point on the density-flow curve represents a traffic state, and the slope of the line connecting the point and the origin describes the aggregated vehicle speed in this state. Assume the mainline traffic is originally in state O (Original state). When the facilitating vehicle decelerates, the vehicles following it also decelerate and accept shorter car-following distances corresponding to the reduced speed. This changes the state of the traffic behind the facilitating vehicle from state O to state C (Cooperative state). The transition in state compacts the mainline vehicles (with higher density) and increases the traffic flow rate, thus providing space for the merging of ramp vehicles. However, the transition from state O to state C also causes a shockwave on the main road, as described by the dashed line connecting state O and state C. The shockwave spreads at the speed defined by the slope of the dashed line and affects the mainline traffic negatively. If the mainline vehicles decelerate too frequently, new shockwaves will be generated before the existing ones dissipate, and the superposition of the shockwaves will cause long-lasting mainline disturbances, eventually leading to traffic breakdowns on the main road [1]. Therefore, the key to CoMC is to balance the efficiency of the mainline and ramp traffic and to ensure that the merging of ramp vehicles is facilitated without breaking mainline stability.

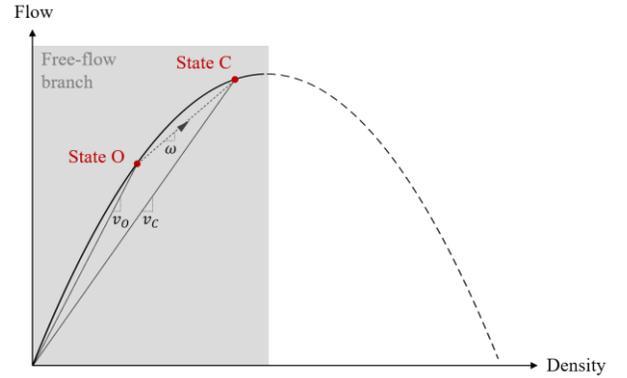

Fig. 2. Generalized fundamental diagram

To this end, the CoMC strategy is dedicated to finding the optimal control scheme that optimizes the overall mainline and ramp efficiency. The control scheme determines the following aspects of the coordination:

- Size of the merging platoon ($n$)
- Movements of the merging platoon ($a$)
- Position at which the facilitating vehicle decelerates ($d$)
- Cooperative speed of the facilitating vehicle ($v_C$)

Note that these aspects are not independent of each other. The optimal control scheme is actually a joint decision on these issues. Moreover, it is assumed that timely communication and precise vehicle control are attainable via the emerging vehicle communication and automation technologies. Specifically, the following assumptions are applied:

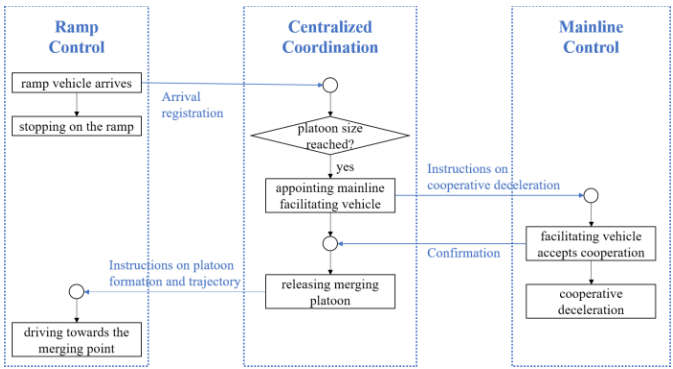

- 100% penetration rate of CAVs in the traffic flow.
- All vehicles are highly-automated corresponding to L4 in [53].
- The control center and relative vehicles are capable of instantaneous communication through vehicle to infrastructure (V2I) technologies.

### B. Analytical Formulation

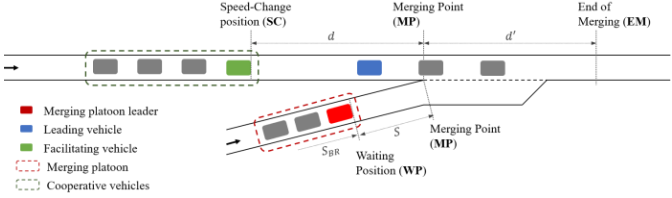

Fig. 3. Definition in a freeway merging area

Fig. 3 shows a hypothetical freeway merging area. The Merging Point (MP) is the position at which the main road and the ramp connect. The merging platoon (marked in red) is formed by stopping the ramp vehicles at the on-ramp Waiting Position (WP), and the mainline cooperation is initiated by the deceleration of the facilitating vehicle at the Speed-Change position (SC). The mainline deceleration may affect several vehicles behind the facilitating vehicle, including the facilitating vehicle itself. These vehicles are called mainline cooperative vehicles (marked in green). The End of Merging (EM) is the position at which the merge influence area ends, as defined in [54]. The entire course of a coordination process, including one gap creation and one platoon formation, is defined as a coordinative merging cycle. CoMC functions through the recurrent implementation of merging cycles.

#### 1) Objective

In order to facilitate the overall merging efficiency, the objective is to minimize the total delay to all vehicles passing through the merging area ($D$), as in (1).

$$min\ D = \left( w_m \cdot \sum_{i=1}^{m} D_{main}^i + w_r \cdot \sum_{j=1}^{n} D_{ramp}^j \right) \times r \quad (1)$$

where $w_m$ and $w_r$ are the weights of the mainline and ramp traffic (this paper uses $w_m = w_r = 1$ for a general case), $m$ is the number of cooperative vehicles in a merging cycle, $n$ is number of ramp vehicles in a merging platoon, $D_{main}^i$ is the delay to the $i$th mainline cooperative vehicle, $D_{ramp}^j$ is the delay to the $j$th ramp vehicle in the platoon, and $r$ is the frequency of merging cycles in number of times per hour.

The delay to a mainline vehicle is defined as the extra amount of time that the vehicle spends on passing the merging area due to the merging of ramp vehicles, namely

$$D_{main}^i = t_{main}^i - t_{main}^0 \quad (2)$$

where $t_{main}^i$ is the actual travel time of the $i$th cooperative vehicle, and $t_{main}^0$ is the original travel time it would take if there were no merging vehicle, with

$$t_{main}^0 = \frac{d + d'}{v_O} \quad (3)$$

where, as in Fig. 3, $d$ is the distance between SC and MP, $d'$ is the distance between MP and EM, and $v_O$ is the speed in original state O. The actual travel time of a mainline vehicle depends on its interaction with the shockwave. As shown in Fig. 4, when the facilitating vehicle decelerates, the mainline traffic changes from state O to state C, generating a shockwave (red dashed line) that propagates along the main road at the speed

$$\omega = \frac{q_C - q_O}{k_C - k_O} \quad (4)$$

where $q_O$, $q_C$ are the flow rates of state O and state C, respectively, and $k_O$, $k_C$ are the corresponding density. We consider the propagation of shockwave ends at EM as vehicles can speed up after leaving the merging bottleneck. A mainline cooperative vehicle, $i$, will initially travel at the speed $v_O$ and decelerate to the speed $v_C$ when encountering the shockwave. Thus, the actual travel time of vehicle $i$ consists of two parts: the travel time in state O ($t_O^i$) and the travel time in state C ($t_C^i$). In state O, we assume that mainline vehicles follow each other at a steady headway of $h_O$ corresponding to the flow rate $q_O$, and set the origin of the time axis to the time point at which the facilitating vehicle passes SC. Then, the $i$th cooperative vehicle will cross SC at the time $(i-1)h_O$ and encounter the shockwave at the time $(i-1)h_O + t_O^i$, meaning that the distance that vehicle $i$ travels in $t_O^i$ seconds equals the distance that the shockwave travels in $(i-1)h_O + t_O^i$ seconds, i.e., $v_O \cdot t_O^i = \omega \cdot [(i-1)h_O + t_O^i]$. This gives $t_O^i$ as

$$t_O^i = \frac{(i-1)\omega h_O}{v_O - \omega} \quad (5)$$

The distance between SC and EM is $d + d'$, which is the sum of the travel distance in state O and the travel distance in state C, i.e., $v_O \cdot t_O^i + v_C \cdot t_C^i = d + d'$. Thus, $t_C^i$ is given as

$$t_C^i = \frac{d + d' - v_O t_O^i}{v_C} \quad (6)$$

Then, the delay to the $i$th cooperative vehicle is

$$D_{main}^i = t_O^i + t_C^i - t_{main}^0 \quad (7)$$

The number of cooperative vehicles in a merging cycle, $m$, depends on the dissipation time of the shockwave $T_{sw}$:

$$T_{sw} = \frac{d + d'}{\omega} \quad (8)$$

For a mainline vehicle $i$ which passes SC at the time $(i-1)h_O$, if it maintains the original speed $v_O$, it will arrive at EM at the time $T_i$, with

$$T_i = (i - 1)h_O + \frac{d + d'}{v_O} \qquad (9)$$

If $T_i < T_{sw}$, vehicle $i$ is a cooperative vehicle because it will encounter the shockwave somewhere between SC and EM; otherwise (i.e., $T_i \geq T_{sw}$), the vehicle will not be affected by the shockwave. When there are exactly $m$ cooperative vehicles in a merging cycle, the $m^{\text{th}}$ vehicle should encounter the shockwave ($T_m < T_{sw}$) and the $m + 1^{\text{th}}$ vehicle should not ($T_{m+1} \geq T_{sw}$), determining $m$ as

$$m = \left\lceil \frac{d + d'}{h_O} \times \left( \frac{1}{\omega} - \frac{1}{v_O} \right) \right\rceil \qquad (10)$$

where $\lceil \cdot \rceil$ represents the nearest upper integer.

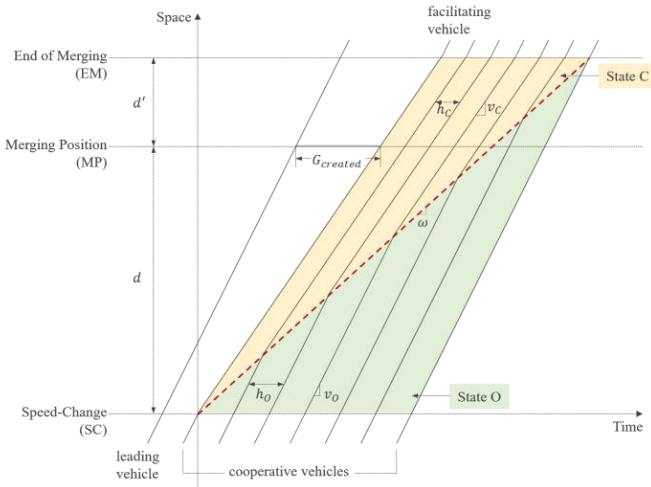

Fig. 4. Spatial-temporal diagram of mainline traffic

The delay to a ramp vehicle $j$ is the difference between its actual travel time ($t_{ramp}^j$) and the original travel time it would take if the main road were empty ($t_{ramp}^0$)

$$D_{ramp}^j = t_{ramp}^j - t_{ramp}^0 \qquad (11)$$

The actual travel time $t_{ramp}^j$ consists of four parts: the time the vehicle spends braking before stopping at WP ($t_{BR}$), the time the vehicle waits at WP ($t_{WT}^j$), the time the vehicle spends accelerating from WP to MP ($t_{ACC}$), and the time the vehicle spends cruising from MP to EM ($t_{CR}$). Note that, we only distinguish a vehicle's position in the platoon for the waiting time at WP, because for the other parts of the trip, vehicles in a platoon behave in exactly the same way.

Assuming that the ramp vehicles arrive at the speed $v_r$ and brake to a stop at a constant braking rate $b$, the braking time ($t_{BR}$) and braking distances ($S_{BR}$) are

$$t_{BR} = \frac{v_r}{b} \qquad (12)$$

$$S_{BR} = \frac{v_r^2}{2b} \qquad (13)$$

Assuming that ramp vehicles arrive in a Poisson distribution $P(\lambda)$ with the arrival rate $\lambda$, the time needed for $x$ ramp vehicles to arrive follows the Gamma distribution $t \sim \Gamma(x, \lambda)$ with the expectation $x/\lambda$. In a merging platoon consisting of $n$ vehicles, the $j^{\text{th}}$ vehicle should wait for $n - j$ vehicles to arrive, so the expected waiting time of the $j^{\text{th}}$ vehicle is

$$t_{WT}^j = \frac{n - j}{\lambda} \qquad (14)$$

According to the requirement of speed, vehicles in the merging platoon should reach the speed $v_C$ and follow each other at an interval equaling the mainline headway in state C ($h_C$) when arriving at MP. Then, the leader of a merging platoon of $n$ vehicles should arrive at MP $n \cdot h_C$ seconds earlier than the mainline facilitating vehicle, so as to satisfy the requirement of arrival time. As CoMC releases the merging platoon at the same time as the facilitating vehicle decelerates at SC, $t_{ACC}$ should be $n \cdot h_C$ seconds less than the facilitating vehicle's travel time from SC to MP (i.e., $d/v_C$), as in (15). During the time $t_{ACC}$, the merging platoon should accelerate from stop to the speed $v_C$, resulting in the required acceleration rate ($a$) and WP position ($S$) as in (16) and (17).

$$t_{ACC} = \frac{d}{v_C} - nh_C \qquad (15)$$

$$a = \frac{v_C}{t_{ACC}} = \frac{v_C^2}{d - nh_C v_C} \qquad (16)$$

$$S = \frac{v_C t_{ACC}}{2} = \frac{d - nh_C v_C}{2} \qquad (17)$$

Here, we adopt the simple assumption of constant ramp acceleration rate. Clearly, more sophisticated trajectory design may benefit energy use and passenger comfort, however, it can be formulated as the task of a lower-level controller and is not the focus of this paper.

After entering the main road, the ramp vehicles should continue travelling at the speed $v_C$ until leaving the merging area, which gives the mainline cruise time $t_{CR}$ as

$$t_{CR} = \frac{d'}{v_C} \qquad (18)$$

The original travel time of a ramp vehicle, $t_{ramp}^0$, is defined as the travel time when the ramp vehicle always travels at the design speed, namely $v_r$ on the ramp and $v_O$ on the main road:

$$t_{ramp}^0 = \frac{S_{BR} + S}{v_r} + \frac{d'}{v_O} \qquad (19)$$

Then, the delay to the $j^{\text{th}}$ ramp vehicle is determined as

$$D_{ramp}^j = t_{BR} + t_{WT}^j + t_{ACC} + t_{CR} - t_{main}^0 \qquad (20)$$

The frequency of coordinative merging cycles, $r$, is related to the merging platoon size $n$. Since a merging cycle is initiated whenever a platoon of $n$ vehicles has formed on the ramp, the expected duration of a cycle ($I$) is equal to the time needed for $n$ ramp vehicles to arrive, as in (21), and the frequency of merging cycles (hourly rate) is the inverse of $I$, as in (22).

$$I = \frac{n}{\lambda} \tag{21}$$

$$r = \frac{3600}{I} = \frac{3600\lambda}{n} \tag{22}$$

### 2) Constraints

The search for objective must be subject to requirements on safety, traffic stability, and vehicle dynamics. Specifically, the followings should apply:

a. The created mainline gap should be no smaller than the space required by the merging platoon, i.e., $G_{create} \geq G_{require}$, with

$$G_{create} = h_O + \frac{d}{v_C} - \frac{d}{v_O} \tag{23}$$

$$G_{require} = (n+1) \cdot h_C \tag{24}$$

b. A new merging cycle can be initiated only when the shockwave caused by the last cooperation has dissipated, i.e., $I \geq T_{sw}$. This can prevent the superposition of disturbances on the main road.

c. The cooperative speed should not fall below the critical speed of mainline traffic, i.e., $v_C \geq v_{crit}$, where $v_{crit}$ is given by the fundamental relationship of mainline traffic flow. This can restrain the negative impacts on the mainline following traffic.

d. The required acceleration of the merging platoon should not exceed the maximum allowable ramp acceleration, i.e., $a \leq a_{max}$.

e. Constraints describing the decision variables by definition

$$\begin{aligned} n &\in \mathbb{N}^+ \\ v_C &< v_O \\ d &> 0 \end{aligned} \tag{25}$$

## III. MODEL SOLUTION

In Section II, the CoMC strategy is formulated as a constrained optimal control problem with three decision variables: (1) the merging platoon size ($n$), (2) the position of SC, described by the distance between SC and MP ($d$), and (3) the cooperative state (state C), described by the speed $v_C$. Table II summarizes the notation and roles of the variables.

TABLE II
NOTATION

| Variable | Symbol | Unit | Role |
|---|---|---|---|
| Merging platoon size | $n$ | veh | decision variable |
| Distance between SC and MP | $d$ | m | decision variable |
| Mainline speed in state C | $v_C$ | m/s | decision variable |
| Mainline flow rate in state C | $q_C$ | veh/h | function of $v_C$ |
| Mainline density in state C | $k_C$ | veh/km | function of $v_C$ |
| Mainline headway in state C | $h_C$ | s | function of $v_C$ |
| Mainline flow rate in state O | $q_O$ | veh/h | input |
| Mainline density in state O | $k_O$ | veh/km | input |
| Mainline speed in state O | $v_O$ | m/s | input |
| Mainline headway in state O | $h_O$ | s | input |
| Shockwave speed | $\omega$ | m/s | function of $v_C$ |
| Number of mainline cooperative vehicles | $m$ | veh | function of $v_C$, $d$ |
| Critical mainline speed | $v_{crit}$ | m/s | input |
| Distance between MP and EM | $d'$ | m | input |
| Ramp arrival rate | $\lambda$ | veh/s | input |
| Ramp arrival speed | $v_r$ | m/s | input |
| Ramp braking rate | $b$ | m/s² | input |
| Maximum ramp acceleration | $a_{max}$ | m/s² | input |

Note that, in the model, we use speed $v$ as the decision variable representing a traffic state, and the other traffic state parameters (flow $q$, density $k$, and headway $h$) are expressed as functions of $v$ according to the fundamental diagram. Theoretically, any form of fundamental diagram would be compatible with CoMC, as the model sets no limits on the macroscopic traffic relationship. In practice, the fundamental relationship can be fitted to field data or derived from a calibrated car-following model. Without loss of generality, we use the fundamental diagram derived from the Wiedemann 99 car-following model as an example:

$$\begin{aligned} s &= CC0 + L + CC1 * v \\ h &= s/v \\ q &= 1/h \\ k &= q/v \end{aligned} \tag{26}$$

where $s$ is the desired spacing distance; $L$ is the vehicle length; $CC0$ and $CC1$ are parameters of the Wiedemann 99 model, representing the standstill distance and the speed-dependent part of $s$, respectively [55].

It is usually difficult to solve such a constrained non-linear non-convex optimization problem. Here, we present an idea to analytically obtain a closely approximated solution. In practice, a heuristic solution to a certain degree of accuracy should also be robust enough.

Based on Section II, the final form of the objective function (with $w_m = w_r = 1$) can be derived as

$$min \ D = \left( \sum_{i=1}^{m} D_{main}^i + \sum_{j=1}^{n} D_{ramp}^j \right) \times r$$

with

$$\sum_{i=1}^{m} D_{main}^i = \frac{m \cdot (v_O - v_C)}{v_C} \\ \times \left[ \frac{d + d'}{v_O} - \frac{(m-1)\omega h_O}{2(v_O - \omega)} \right] \tag{27}$$

$$\sum_{j=1}^{n} D_{ramp}^j = n \times \left( \frac{v_r}{2b} + \frac{d + d'}{v_C} - n h_C - \frac{d - n h_C v_C}{2 v_r} \right. \\ \left. - \frac{d'}{v_O} + \frac{n-1}{2\lambda} \right)$$

$$r = \frac{3600\lambda}{n}$$

If we relax the integer requirement in (10) and approximate $m$ to $m \approx \frac{d+d'}{h_O} \times \left( \frac{1}{\omega} - \frac{1}{v_O} \right)$, the objective function is transformed into a quadratic function of the variable $d$, in the form of

$$min\ D(d) = A \cdot d^2 + B \cdot d + C$$
with

$$A = \frac{3600\lambda(v_o - v_C)(v_o - \omega)}{2n \cdot v_o{}^2 \cdot v_C \cdot \omega \cdot h_o}$$

$$B = \frac{3600\lambda(v_o - v_C)[2d'(v_o - \omega) + v_o \cdot \omega \cdot h_o]}{2n \cdot v_o{}^2 \cdot v_C \cdot \omega \cdot h_o}$$
$$+ 3600\lambda\left(\frac{1}{v_C} - \frac{1}{2v_r}\right)$$

(28)

Because all parameters are positive and $\omega < v_C < v_o$ holds according to the fundamental diagram, $A > 0$ and the parabola of the objective function points upwards with respect to $d$. The symmetrical axis is at the position $d = -B/2A$. With any given pair of $n$ and $v_C$, the feasible range of $d$ is limited by the constraints, where constraints a and d define the lower bound of $d$ under a given pair of $n$ and $v_C$ ($d_{lb}|(n, v_C)$), and constraints b defines the upper bound ($d_{ub}|(n, v_C)$), as in (29)

$$d \geq d_{lb}|(n, v_C)$$
$$= max\begin{pmatrix} \dfrac{v_o v_C}{v_o - v_C}[(n+1)h_C - h_O] \\ \dfrac{v_C{}^2}{a_{max}} + nh_C v_C \end{pmatrix}$$
$$d \leq d_{ub}|(n, v_C) = \frac{n\omega}{\lambda} - d'$$

(29)

Then, the optimal value of $d$ under a given pair of $n$ and $v_C$ ($d_{opt}|(n, v_C)$) can be obtained by comparing the position of the symmetrical axis to the upper and lower bounds of $d$, namely

$$d_{opt}|(n, v_C)$$
$$= \begin{cases} d_{lb}|(n, v_C), & if\ -\dfrac{B}{2A} \leq d_{lb}|(n, v_C) \\ -\dfrac{B}{2A}, & if\ d_{lb}|(n, v_C) < -\dfrac{B}{2A} < d_{ub}|(n, v_C) \\ d_{ub}|(n, v_C), & if\ -\dfrac{B}{2A} \geq d_{ub}|(n, v_C) \end{cases}$$

(30)

The optimal objective value under a given pair of $n$ and $v_C$, $D_{opt}|(n, v_C)$, is obtained by substituting $d_{opt}|(n, v_C)$ into the objective function, i.e., $D_{opt}|(n, v_C) = D\left(d_{opt}|(n, v_C)\right)$.

Note that for any given pair of $n$ and $v_C$, the optimal problem has solutions only when $d_{lb}|(n, v_C) \leq d_{ub}|(n, v_C)$ holds. This defines the feasible range of $n$ under a given $v_C$. Considering that $n$ must be a positive integer, we can check all possible values of $n$ and find the optimal $n$ that produces the minimal value of $D$ for that $v_C$, i.e.,

$$D_{opt}|v_C = \min_{n=k, k+1, k+2, \dots} \left(D_{opt}|(n, v_C)\right)$$
$$n_{opt}|v_C = \arg\min_{n=k, k+1, k+2, \dots} \left(D_{opt}|(n, v_C)\right)$$

(31)

where $k, k + 1, k + 2, \dots$ stand for the feasible range of $n$ under a given $v_C$, and $n_{opt}|v_C$ and $D_{opt}|v_C$ are the optimal values of $n$ and objective $D$ under that $v_C$, respectively.

So far, we have presented a way to determine the optimal values of $d$ and $n$ given the value of $v_C$, so the relationship between $v_C$ and the optimal $D$ value that is achieved under the particular $v_C$ ($D_{opt}|v_C$) can be established, see Fig. 5 as an example. The optimal solution locates at the point where $D_{opt}|v_C$ is minimal (the red dot in Fig. 5).

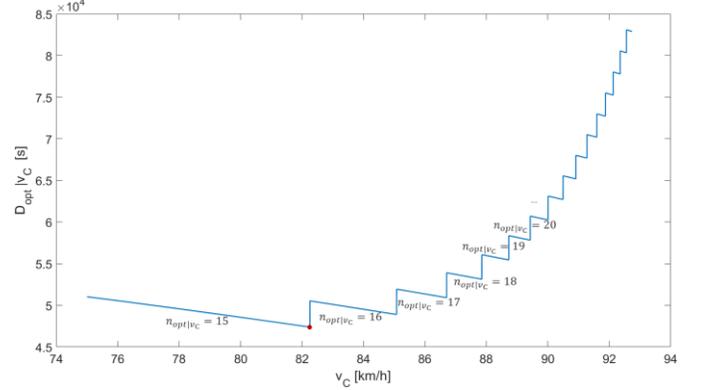

Fig. 5. $v_C$-$D_{opt}|v_C$ relationship (with mainline volume 1800 veh/h, ramp volume 500 veh/h, and parameter values in Table III)

As shown in Fig. 5, $D_{opt}|v_C$ monotonically decreases as $v_C$ increases, but when $v_C$ reaches certain levels (i.e., breakpoints of the curve in Fig. 5), the value of $D_{opt}|v_C$ will suddenly increase. This is because, as $v_C$ increases to the breakpoints, the constraint $d_{lb}|(n, v_C) \leq d_{ub}|(n, v_C)$ will fail with the current value of $n_{opt}|v_C$, so $n_{opt}|v_C$ must increase by 1. Therefore, the optimal solution must appear at one of the breakpoints. By comparing the values of $D$ at the breakpoints, the optimal solution (marked as $d^*$, $n^*$, and $v_C{}^*$) can be obtained as

$$v_C{}^* = \arg\min_{v_C} D_{opt}|v_C$$
$$n^* = n_{opt}|v_C{}^*$$
$$d^* = d_{opt}|(n^*, v_C{}^*)$$

(32)

## IV. PERFORMANCE ANALYSIS

The analytical formulation in Section II is derived from the theoretical traffic flow models. In this section, we conduct an illustrative case study to verify the efficiency of CoMC in the more complicated simulation environment.

### A. Simulation Design

The simulation platform is composed of the microsimulation tool VISSIM version 11.8 and scripts coded in Python version 3.6 and C++ version 2017. VISSIM provides the basic simulation environment, including the road network, traffic demand generation, lower-level vehicle dynamics, and raw data record. The upper-level coordination of CoMC is compiled in Python and integrated into VISSIM through the COM interface. The cooperative behaviors of the mainline facilitating vehicle and the merging platoon leader are controlled by external driving models written in C++ and called via the DLL interface of VISSIM.

The simulated freeway extends 2000 meters upstream and 500 meters downstream from the merging area, covering the merge influence area defined in HCM [54]. A 700-meter-long on-ramp connects to the main road via a 240-meter-long acceleration lane. For each study scenario, 10 simulation runs with different random seeds are carried out. Each run lasts 7200 simulation seconds. The reported results are aggregated over the 10 replications of each study scenario.

Fig. 6 shows the maximum on-ramp volume that can be accommodated by CoMC with respect to the mainline flow rate for the input parameters in Table III. The range of flow covers most prevailing field-observed demand scenarios, indicating the ability of CoMC to address real-world merging problems.

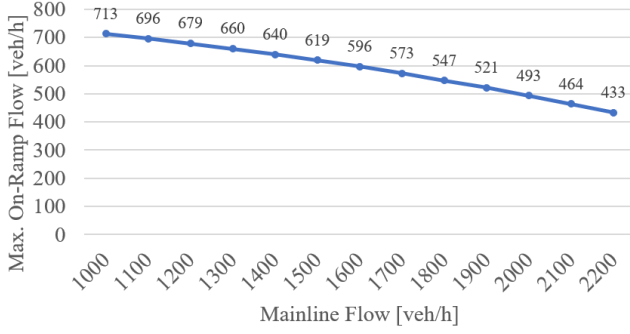

Fig. 6. Maximum on-ramp flow with respect to mainline flow

<div style="text-align:center">

TABLE III
INPUT PARAMETERS

</div>

| Parameter | Value | Unit | Source |
|-----------|-------|------|--------|
| $v_o$ | 120 | km/h | [54] |
| $v_r$ | 60 | km/h | [54] |
| $d'$ | 457.2 | m | [54] |
| $v_{crit}$ | 75 | km/h | [56] |
| $b$ | 2.75 | m/s$^2$ | [57] |
| $a_{max}$ | 2.75 | m/s$^2$ | [57] |
| $CC0$ | 1.5 | m | [58] |
| $CC1$ | 0.9 | s | [58] |
| $L$ | 4.37 | m | [58] |

As the traffic is usually self-sustained under low traffic demand, the main purpose of CoMC is to promote merging and prevent congestions under high traffic volume conditions. In this case study, we focus on the high demand scenarios and consider two levels of mainline flow (1600 and 1800 veh/h) and three levels of on-ramp flow (300, 400, and 500 veh/h), resulting in a total of six demand scenarios, as shown in Table IV. Note that, the demand values in Table IV only serve as reference values; in the simulation, vehicles are generated at random headways, resulting in fluctuating flow rate around the reference values. For each demand scenario, a CoMC-controlled case is developed and compared to a base case where no upper-level control is applied. The base and CoMC cases use exactly the same assumptions and models with the only difference as the presence/absence of CoMC. According to the CoMC control plan in Table IV, the required values of $v_C$, $n$, and $d$ are reasonable in all scenarios, showing that CoMC is compatible with real-world implementation.

<div style="text-align:center">

TABLE IV
DEMAND SCENARIO AND CONTROL PLAN

</div>

|  | 1A | 1B | 1C | 2A | 2B | 2C |  |
|---|-----|-----|-----|-----|-----|-----|---|
| $q_{main}$ | 1600 | 1600 | 1600 | 1800 | 1800 | 1800 | veh/h |
| $q_{ramp}$ | 300 | 400 | 500 | 300 | 400 | 500 | veh/h |
| $v_C$ | 96.67 | 89.80 | 83.53 | 99.61 | 88.16 | 82.25 | km/h |
| $d$ | 624 | 794 | 1062 | 911 | 847 | 1266 | m |
| $n$ | 4 | 7 | 12 | 5 | 8 | 15 | veh |

### B. Results and Discussion

In order to visualize the simulated traffic conditions, we record the speed and position of each vehicle in the road segment between 1500 meter upstream and 500 meters downstream of the merging point at the start of each simulation second and plot vehicle trajectories, as shown in Fig. 7, where black lines indicating the mainline vehicles, and the red lines the merging vehicles. As it shows, the expected coordination phenomena, such as the formation of merging platoons, the cooperative deceleration of mainline vehicles, and the propagation and dissipation of disturbances are observed in the simulation environment. Under relatively low traffic volumes (e.g., 1A and 1B), the disturbances induced by the merging vehicles are usually quickly eliminated in both base and CoMC cases. Under high traffic volumes (e.g., 1C, 2B, and 2C), the disturbances induced by the merging traffic may accumulate and eventually trigger traffic breakdowns in the base cases, whereas the periodic coordination of CoMC can well collect and accommodate the disturbances on the main road.

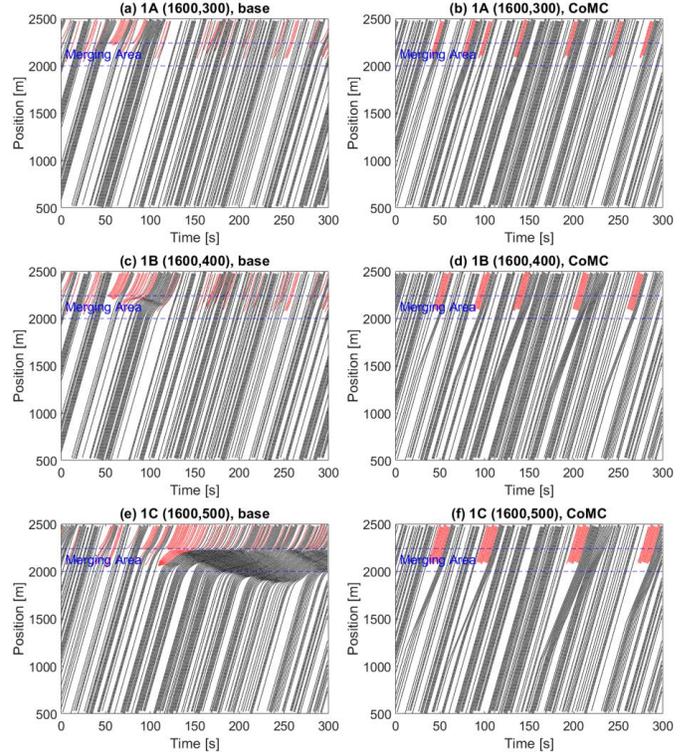

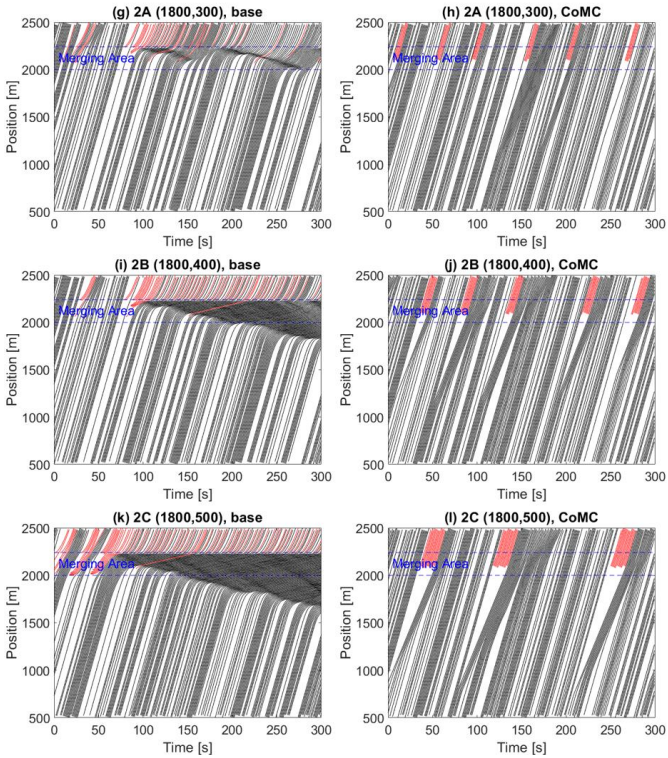

Fig. 7. Five-minutes trajectory plot

Table V, Table VI and Fig. 8 present the travel time and delay results. The travel time results are measured for the entire vehicle trips, excluding the first and last 100 meters to switch off border effects. Delay is defined as the difference between the measured travel time and the theoretical ideal travel time, with the ideal travel time referring to the minimum driving time determined by the length and design speed of each road link. In addition, we collect the average cross-section speed at every 100 meters from 500 meters to 2500 meters along the mainline freeway at a time frequency of five simulation minutes and report the aggregated speed results in Fig. 9 to reflect the prevailing traffic conditions and the onset of congestions.

According to the results, CoMC may increase the ramp travel time and delay, as in scenarios 1A and 1B, because in these scenarios where the traffic performs well even without control, it is unnecessary to pause the merging vehicles on the ramp. In the other scenarios with higher traffic volumes, CoMC can reduce both the mainline and ramp delay to different extents. A large improvement in the ramp efficiency is observed in the scenarios 2A and 2B, where in the base cases, the ramp vehicles can hardly merge in time due to the high mainline volume, and this problem is improved by CoMC through the proactively created on-demand gaps. The most remarkable efficiency gain is observed in the most critical scenario 2C, where the overall travel time and delay are reduced by 67.3% and 91.6%, respectively, with CoMC. This is resulted from CoMC's ability to stabilize traffic and prevent recurrent congestions at ramp merging. As shown in Fig. 9e, i, and k, under high traffic volumes, the intensive merging of ramp vehicles may trigger traffic breakdowns in the uncontrolled base cases, and the congestions may persist and even spread upstream along the main road in the most critical case. When CoMC is applied, the periodic coordination can ensure both the timely merging of ramp vehicles and the recovery of mainline stability, thereby guaranteeing a fluent operation of traffic even under the high traffic volume conditions. As a result of the prevention of traffic breakdowns and capacity drop phenomena, the case study show that CoMC can increase the overall throughput of the merging area from 2128.5 veh/h to 2269.5 veh/h (approximately 6.6%) in the critical 2C scenario.

TABLE V
TRAVEL TIME RESULTS

| | | Mainline travel time (s) | Ramp travel time (s) | Overall travel time (s) |
|---|---|---|---|---|
| 1A | base | 77.22 | 70.93 | 76.23 |
| (1600,300) | CoMC | 77.01 | 76.73 | 76.97 |
| 1B | base | 78.72 | 85.60 | 80.09 |
| (1600,400) | CoMC | 77.89 | 87.01 | 79.71 |
| 1C | base | 103.76 | 104.78 | 104.00 |
| (1600,500) | CoMC | 79.42 | 100.48 | 84.42 |
| 2A | base | 80.31 | 111.80 | 84.76 |
| (1800,300) | CoMC | 77.16 | 83.15 | 78.01 |
| 2B | base | 111.47 | 127.03 | 114.31 |
| (1800,400) | CoMC | 78.81 | 92.08 | 81.22 |
| 2C | base | 311.92 | 128.01 | 269.67 |
| (1800,500) | CoMC | 81.47 | 112.77 | 88.23 |

TABLE VI
DELAY RESULTS

| | | Mainline delay (s) | Ramp delay (s) | Overall delay (s) |
|---|---|---|---|---|
| 1A | base | 1.02 | 15.73 | 3.34 |
| (1600,300) | CoMC | 0.81 | 21.53 | 4.08 |
| 1B | base | 2.52 | 30.40 | 8.09 |
| (1600,400) | CoMC | 1.69 | 31.81 | 7.71 |
| 1C | base | 27.56 | 49.58 | 32.83 |
| (1600,500) | CoMC | 3.22 | 45.28 | 13.20 |
| 2A | base | 4.11 | 56.60 | 11.53 |
| (1800,300) | CoMC | 0.96 | 27.95 | 4.81 |
| 2B | base | 35.27 | 71.83 | 41.95 |
| (1800,400) | CoMC | 2.61 | 36.88 | 8.83 |
| 2C | base | 235.72 | 72.81 | 198.29 |
| (1800,500) | CoMC | 5.27 | 57.57 | 16.57 |

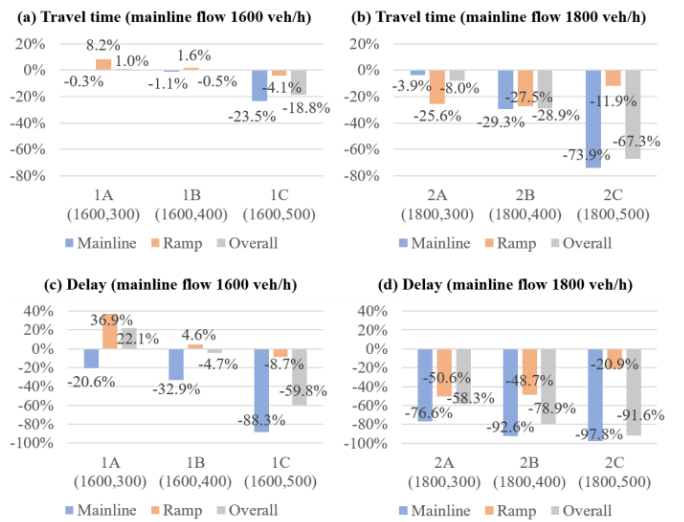

Fig. 8. Changes in travel time and delay with the implementation of CoMC

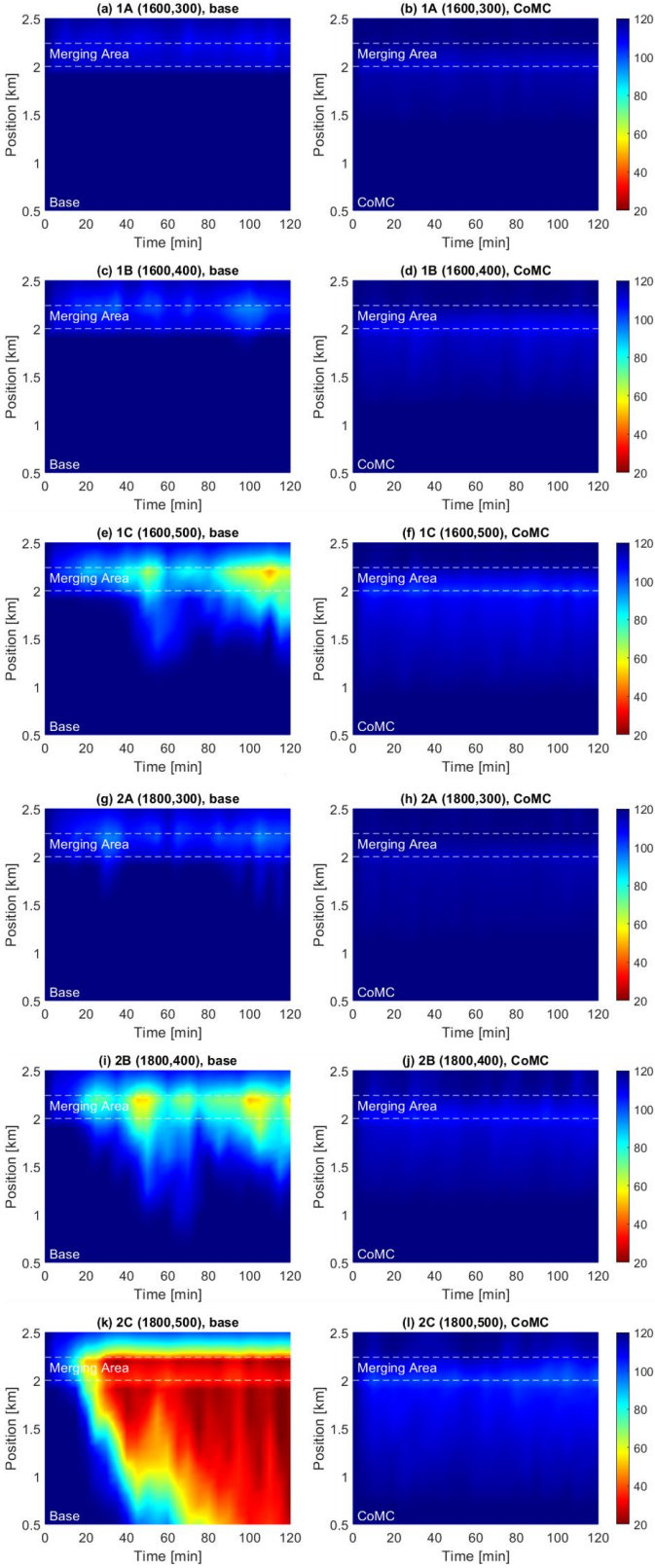

Fig. 9. Speed contour

As revealed in the case study, the benefits of CoMC are more remarkable under higher traffic volume. In the cases where the traffic volume is low and external coordination actually not needed, CoMC may cause extra delays to the ramp traffic. Therefore, we recommend introducing a threshold in terms of traffic volumes, at which the CoMC strategy would be activated in practice. This activation threshold should be determined on a 'case by case' basis in the light of the specific conditions of a merging area, as traffic conditions may vary across different sites.

Further, as the real-word traffic state may vary over time, the CoMC strategy should be implemented in a dynamic manner in practice. Specifically, the traffic state parameters (i.e., aggregated flow rate and speed) should be estimated at regular intervals, and the decisions of CoMC (i.e., activate or not, if activate, values of the control parameters $n$, $v_C$, and $d$) should be updated recurrently according to the latest state estimations. The update interval can vary at different sites or times of day. For example, when the traffic condition is relatively stable (e.g., during night-time), the coordination can be updated at a longer interval (e.g., 30 minutes or longer), whereas in the situations with dramatic changes in the traffic state (e.g., transition periods between the non-peak and peak hours), the traffic condition should be actively monitored, and the CoMC decisions should be updated at a higher frequency (e.g., every 5 to 10 minutes). The dynamic implementation method can enhance the adaptivity of the coordination and ensure that the real-time variations in traffic conditions are accommodated in a timely manner.

## V. CONCLUSION AND FUTURE WORK

In this paper, we present a novel CAV merging strategy, called CoMC, to coordinate traffic flows in freeway on-ramp bottlenecks and facilitate ramp merging operation. The strategy proactively creates on-demand gaps on the mainline freeway and guides the ramp vehicles into the created gaps in the form of platoons. The proposed strategy is formulated as an optimization problem that minimizes total vehicle delay based on the macroscopic and microscopic traffic flow models. The analytical model determines the optimal control plan with respect to the traffic conditions, including the cooperative merging speed, the merging platoon size, and the speed-change position of the mainline facilitating vehicle. The efficiency of CoMC under various demand scenarios is demonstrated in a case study conducted on a microscopic simulation platform. The results show that the proposed CoMC coordination is successfully achieved in the simulation environment and can substantially improve the overall traffic efficiency at on-ramp merging, especially under high traffic volume conditions. In the most critical conditions, recurrent traffic congestions are prevented, and merging throughput is increased. In comparison with the existing CAV merging strategies, the novelty of CoMC resides on the following aspects: (1) it focuses on the upper-level coordination of two streams of traffic (instead of individual vehicles) for the flow-level efficiency gains; (2) it presents an innovative idea combining gap-creation and platoon-merging, especially the detailed discussion on platoon merging is unique; (3) macroscopic traffic flow models are integrated, allowing for an explicit consideration on traffic flow stability.

The current work has some limitations that should be addressed in the future. First, the current CoMC system requires a 100% penetration rate of CAVs, which is not in line with the prediction of the near future situation, where CAVs and HDVs may coexist in the freeway network. In fact, the CoMC strategy has the potential to be extended for a mixed traffic condition, because it only requires the CAV capabilities from the facilitating vehicle and the platoon leader, whereas the other vehicles can be maneuvered by human drivers through the normal car-following rules. However, further investigations are required to assign the vehicles' roles in the coordination and to accommodate the uncertainties introduced by the HDVs. Further, there is a potential to enhance the coordination in multilane freeway layouts by, for example, including a proactive centralized control of courtesy lane-changes. However, further considerations are required to ensure that the cooperation will not cause excessive interference to the upstream mainline traffic. These extended strategies are the main focuses of our on-going research.

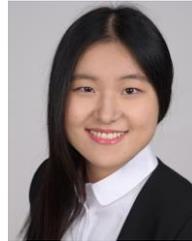

**Jie Zhu** received the B.Sc. degree in transportation engineering from Tongji University, Shanghai, China, in 2014, and the M.Sc. degree in civil engineering from Technical University of Munich, Germany, in 2017. She is currently a Ph.D. candidate at Chalmers University of Technology, Gothenburg, Sweden. Her research is focused on the operation and control of emerging intelligent vehicles.

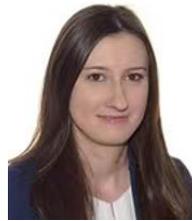

**Ivana Tasic** is an assistant professor at Chalmers University of Technology. Her main research interests are urban multimodal transportation systems, innovative transportation technologies, and multidisciplinary education. She holds a B.Sc. degree in Transportation engineering from the University of Belgrade, Serbia, and a Ph.D. degree in Civil Engineering/Transportation from the University of Utah. She is trained as a multidisciplinary researcher in transportation infrastructure and urban planning and has been supporting multiple professional engineering organizations through active membership and service.

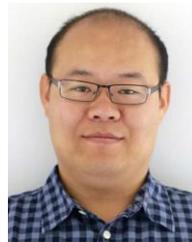

**Xiaobo Qu** (Member, IEEE) received the B.Eng. degree from Jilin University, Changchun, China, the M.Eng. degree from Tsinghua University, Beijing, China, and the Ph.D. degree from the National University of Singapore, Singapore. He is currently a Professor and the Chair of the Urban Mobility Systems of Chalmers University of Technology, Gothenburg, Sweden. His research is focused on integrating emerging technologies into urban transport systems. He is an elected member of Academia Europaea – the Academy of Europe.

TABLE I
COMPARISON TO THE STATE-OF-THE-ARTS

| Strategy | CAV penetration | Freeway layout | Control strategy | | Optimal merging platoon [a] | Macroscopic traffic models [b] |
|---|---|---|---|---|---|---|
| | | | Upper-level | Lower-level | | |
| CoMC | 100% | single-lane | on-demand gap creation & platoon merge | - | ✓ | ✓ |
| Cao, et al. [22] | 100% | single-lane | - | trajectory planning | - | - |
| Zhou, et al. [23] | 100% | single-lane | - | trajectory planning | - | - |
| Zhou, et al. [24] | 100% | single-lane | - | trajectory planning | - | - |
| Fukuyama [27] | 100% | single-lane | - | trajectory planning | - | - |
| Karimi, et al. [28] | mixed | single-lane | - | trajectory planning | - | - |
| Letter, et al. [29] | 100% | single-lane | - | trajectory planning | - | - |
| Xie, et al. [30] | 100% | single-lane | - | trajectory planning | - | - |
| Hu, et al. [31] | 100% | multi-lane | - | trajectory planning | - | - |
| Omidvar, et al. [32] | mixed | single-lane | - | trajectory planning | - | - |
| Ito, et al. [33] | mixed | single-lane | - | trajectory planning | - | - |
| Mu, et al. [34] | 100% & mixed | single-lane | - | trajectory planning | - | - |
| Ntousakis, et al. [35] | 100% | single-lane | - | trajectory planning | - | - |
| Rios-Torres et al. [36] | 100% | single-lane | - | trajectory planning | - | - |
| Sonbolestan, et al. [38] | 100% | single-lane | - | trajectory planning | - | - |
| Karbalaieali, et al. [39] | 100% | multi-lane | - | alternative action | - | - |
| Ding, et al. [40] | 100% | single-lane | local merging sequence | trajectory planning | - | - |
| Jing, et al. [41] | 100% | single-lane | local merging sequence | trajectory planning | - | - |
| Xu, et al. [42] | 100% | single-lane | local merging sequence | trajectory planning | - | - |
| Chen, et al. [43] | 100% | single-lane | local merging gap | trajectory planning | - | - |
| Sun, et al. [44] | mixed | single-lane | local merging gap | trajectory planning | - | - |
| Scarinci, et al. [45] | 100% | single-lane | periodic gap creation | - | - | ✓ |
| Chen, et al. [46] | mixed | single-lane | periodic gap creation & batch merge | - | - | ✓ |

[a] indicating whether the strategy considers platoon merging and determines the optimal platoon size.
[b] indicating whether the strategy incorporates macroscopic traffic flow models and controls the fundamental state of the traffic flow.